\documentclass[a4paper]{article}
\usepackage{url,xcolor}
\usepackage{authblk}
\usepackage{graphicx}
\usepackage{float,afterpage}
\usepackage[margin=3cm]{geometry}
\usepackage[numbers]{natbib}
\usepackage{amsmath,amssymb,amsfonts}
\title{Mapping, modeling, and reprogramming cell-fate decision making systems}

\author[1,2,3,*]{Lucy Ham}
\author[2,*]{Taylor E. Woodward}
\author[1,4]{Megan A. Coomer}
\author[1,2]{Michael P.H. Stumpf}

\affil[1]{School of BioSciences, University of Melbourne, Parkville, Australia}
\affil[2]{School of Mathematics and Statistics, University of Melbourne, Parkville, Australia}
\affil[3]{ARC Centre of Excellence for the Mathematical Analysis of Cellular Systems, University of Melbourne, Parkville, Australia}
\affil[4]{Cell Bauhaus, Melbourne, Australia}
\affil[*]{These authors contributed equally}
\begin{document}

\maketitle

\begin{abstract}
Many cellular  processes involve information processing and decision making. We can probe these processes at increasing molecular detail. The analysis of heterogeneous data remains a challenge that requires new ways of thinking about cells in quantitative, predictive, and mechanistic ways. We discuss the role of mathematical models in the context of cell-fate decision making systems across the tree of life. Complex multi-cellular organisms have been a particular focus, but single celled organisms also have to sense and respond to their environment. We center our discussion around the idea of design principles which we can learn from observations and modeling, and exploit in order to (re)-design  or guide cellular behavior.
\end{abstract}



\section{Introduction}
All cells in our bodies derive from a single ancestor, the fertilized egg cell. A process of division and differentiation gives rise to all 35-40 Trillion cells~\cite{Hatton:2023aa}. How the correct types of cells appear in their correct numbers, at the right place, and at the right time remains a riddle central to developmental and stem-cell biology. It is an instructive riddle which goes to the core of biology. The genome alone does not suffice to make an organism; the cellular machinery provides the physiological context and integrates external stimuli which have all come together in orchestrating which genes are expressed when, where, and for how long~\cite{rizvi2017,macarthur2022}. 
\par

Cell-fate decisions are generally considered within the context of multicellular organisms, where stem or progenitor cells undergo a series of commitment steps to adopt specific downstream fates. But it is beneficial to take a broader view and also consider decision making systems in single-celled organisms, including bacteria. {\em Bacillus subtilis} is an excellent example of a bacterium that ``makes" decisions akin to those we see in multi-cellular organisms~\cite{Kuchina:2011gh,Iwanska2024}. 
\emph{B. subtilis} can undergo sporulation, producing semi-inert spores that act as carriers of genetic information, surviving for extended periods and reactivating into normal bacterial cells when conditions improve. Alternatively, B. subtilis can form biofilms~\cite{Dragos2021}, which are organised communities with both local and large-scale spatial coherence and have even been described as possessing ``embryo-like features”~\cite{Futo2021}.
 Not all strains of {\em B. subtilis} can exhibit sporulation or biofilm formation, which we must assume reflects  different genomic contents and contexts among the different strains subsumed under {\em B. subtilis}. 
\par
The conservation of biological processes across vast evolutionary distances is well-documented in the literature~\cite{Stadler:2021jj,Foley2023}. Jacques Monod was only slightly -- if at all -- facetious when saying: ``What is true for {\em E. coli} is true for the elephant", highlighting that fundamental mechanisms can be shared across diverse life forms. Key aspects of transcriptional regulation in mammalian systems, for example, were elucidated through foundational studies like Monod’s work on the {\em lac} operon in {\em Escherichia coli}.
While the outcomes of cell-fate decision making systems in microbial organisms may lack the intricacy and variety of multi-cellular systems, they can offer us insights into fundamental aspects of cell-fate decision making systems and processes~\cite{Futo2021}. But microbial systems are also important in their own right and being able to affect or reprogram their behavior holds great potential for biotechnology, synthetic and engineering biology, and the management of bacterial and fungal pathogens.  

\par
We now have access to a bewildering array of experimental assays that allow us to study aspects of cell-fate decision making processes and systems. But data alone does not equate to knowledge, and even the most cutting edge experimental assays and technologies only deliver a partial view of the developmental processes involved in cell-fate decisions.  
Currently, single-cell data is abundant for mRNA, and a wide range of tools and techniques have emerged to analyze this information, as detailed in excellent reviews elsewhere \cite{Baysoy:2023,Heumos:2023,Gorin:2022,Moses:2022}. Other molecular states are also becoming accessible at single-cell resolution. For example, single-cell ATAC-seq (Assay for Transposase-Accessible Chromatin) reveals genome accessibility to regulatory machinery. Although single-cell proteomics is available, it remains limited to quantifying a small subset of proteins or post-translationally modified proteins. Single-cell metabolic analysis is similarly in an early stage of development~\cite{Ahl2020,Rappez2021,Hrovatin2023}. 
\par
Even though we expect that these single-cell techniques will develop rapidly, there are two impediments to the experimental analysis of cell-fate decision making systems that will continue to persist in the short term: first, we lack the ability to collect several data modalities simultaneously in the same cell; second, we cannot at scale monitor cells over time. This is due to the destructive nature of most assays and the limited resolution of non-destructive (observational or imaging-based) technologies. This is a problem for our ability to integrate and draw unified inferences from multi-omic data.  Though we have made progress in integrating bulk data, single-cell analysis presents unique challenges for combining epigenomic, metabolomic, transcriptomic, and proteomic data. This is also a challenge for state-of-the-art machine learning (ML) or artificial intelligence (AI) models.   These approaches rely on the detection and analysis of statistical features in large complex data sets, but without simultaneous mRNA and metabolomic data from the same cells, crucial dependencies are lost, reducing the models' predictive power. There are computational strategies to address some of these challenges, which we will discuss below.

\par
Mechanistic modeling, especially when coupled with ML/AI, offers valuable alternatives for understanding cell-fate decision making processes~\cite{Casey:2020bq,brun-usan2020,Babtie:2017ix}. Here by ``mechanistic models," we mean mathematical and computational representations of our biological understanding, hypotheses, and assumptions. These models are (i) quantitative; (ii) predictive; and (iii) correspond to hypotheses that can be tested, evaluated, and refined in light of new data and information. In contrast, purely ML/AI models, though predictive, often lack mechanistic insights. While magnificent resources like AlphaFold~\cite{jumper2021}, provide powerful descriptive insights, extracting actionable knowledge from them still requires additional interpretation and refinement~\cite{david2022, Madsen2023}.  

\par 
This review focuses on extracting knowledge from experimental data. We argue that mechanistic models are essential tools for interpreting cell-fate decision making systems. Mechanistic models not only serve as conceptual frameworks, but also enable predictive analyses, making them invaluable for deciphering data in a meaningful way. In our view, an effective analysis of cell-fate systems requires a balance: a purely conceptual but non-predictive approach is as limited as a purely predictive model with no functional or mechanistic understanding.
One potentially positive outcome of centering mechanistic models at the center of cell-fate research could be a shift away from descriptive data collection to ``discriminatory data" collection,  meaning data that directly tests and informs our hypotheses and models.  Large-scale Cell Atlas projects provide a rich backdrop, allowing for more focused and detailed investigations into the molecular mechanisms driving cell fates.
\par
We begin by establishing foundational concepts in cell-fate decision making processes, with a primary focus on mammalian systems. In this domain, experimental methodologies have progressed enormously and we are now able to map biological processes at single-cell resolution, especially at the transcriptional level.
While protein, epigenetic, lipid, and metabolic analyses are still developing, they are expected to reach comparable precision soon. Due to their larger size and advanced study capabilities, mammalian systems offer insights that may also help us understand cell-fate processes in simpler organisms, such as microbes.

Following this, we explore the critical role of modeling in interpreting the wealth of cell-fate data. Modeling serves two primary purposes: (1) tackling the inverse problem in cell-fate decision making, that is, inferring the underlying mechanisms from observed data, and (2) developing frameworks, which we term “CellMaps” (following the precedent set by Sydney Brenner), to potentially reprogram cellular behavior. While this capability is still largely aspirational, advances in synthetic and engineering biology bring us close to achieving it.

\par
We then shift to constructing realistic CellMaps and whole-cell models, focusing on microbial systems where these approaches have already seen success, particularly in synthetic and engineering biology.
 The aim is to generate new genetic constructs in biological organisms, or to design and construct new strains that are better capable of performing specific functions, from biosensing to bioprocessing. 
Bacteria and single-celled eukaryotes (archaea are much less well studied), with their simpler genomes and molecular networks, provide a more feasible starting point for whole-cell modeling than multicellular organisms. We outline some of the current approaches, early successes, and promising developments in modeling aimed at supporting synthetic/engineering biology and biotechnology. Finally, we emphasize the importance of comparative biology. The lessons we learn in the context of microbial lifeforms have the potential to enhance our understanding of metazoan and mammalian systems, and vice versa. The feedback between these domains will be invaluable for advancing both our theoretical and practical knowledge of cell-fate decision making across the tree of life.
\begin{figure}[h]
\includegraphics[width=\textwidth]{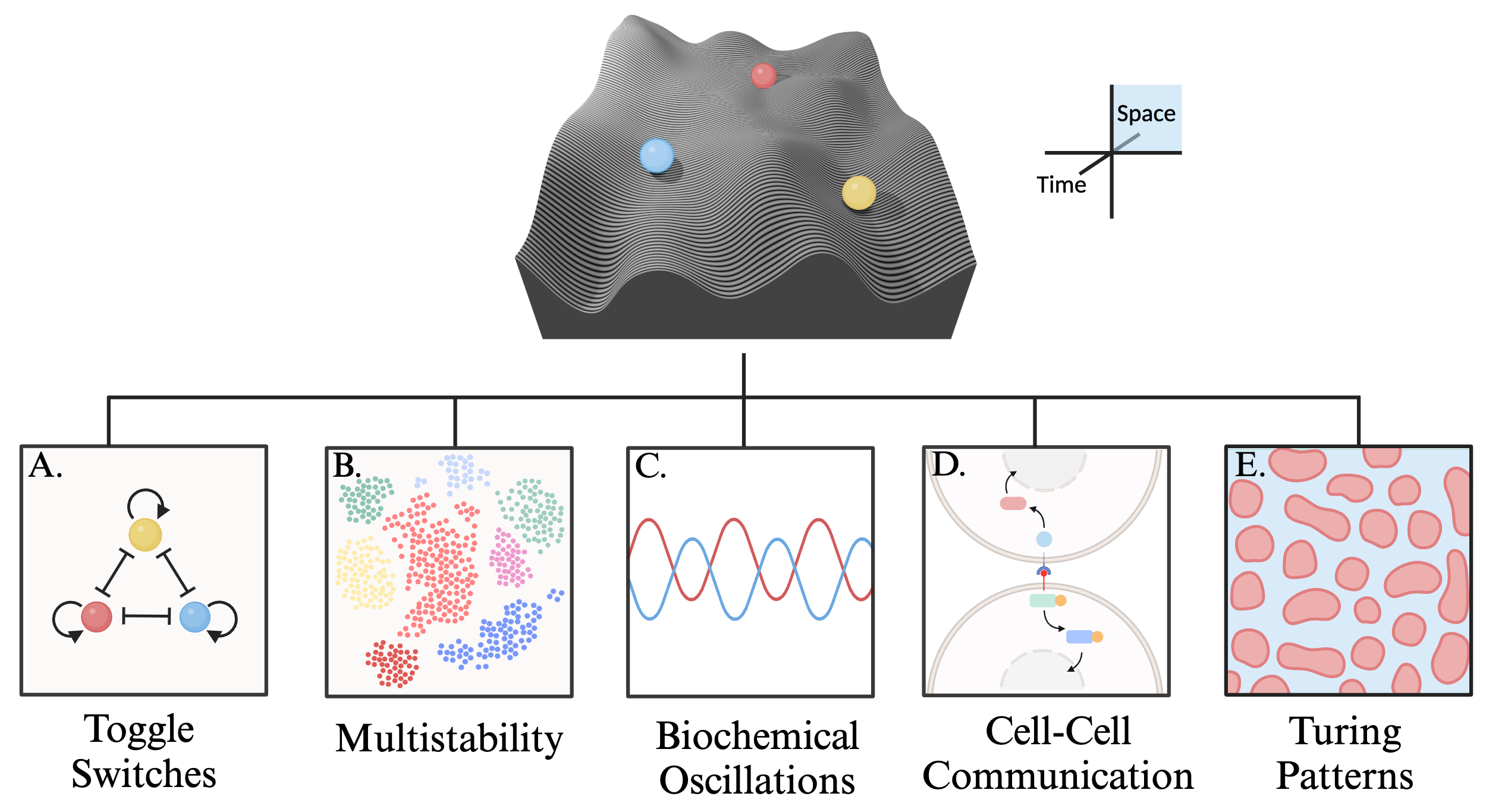}
\caption{ 
Illustration of key design principles in cell-fate decision making systems over space and time. (A) Toggle Switches enable cells to adopt and stably maintain specific states through mutually inhibitory feedback loops. (B) Multistability is the ability of cells to attain multiple stable states, allowing for cellular diversity within a genetically identical population. (C) Oscillatory dynamics in gene expression provide timing and rhythmic control in developmental processes. (D) Cell communication is crucial for robust cell-fate control.  Cells exchange signals with neighbors, enabling coordinated behavior and stabilization of gene expression patterns across tissues. (E)  Spatial patterns emerge from reaction-diffusion processes, allowing cells to self-organize into distinct regions, as observed in animal coat patterns and tissue structures.
This figure was created in BioRender.com.
}
\label{fig1:Design_principles}
\end{figure}


\section{Cell-Fate Decision Making Systems}

We begin by outlining the role of design principles in cell-fate decision making, followed by working definitions of cell states, types, and fates. We then examine exemplar cell-fate decision making systems, first qualitatively and then quantitatively.
\par
There is still some debate about what constitutes a cellular state, a cell-fate, or a cell type \cite{Clevers:2017fu,Rafelski2023}. 
While various biological definitions have been proposed, greater clarity is also needed from a theoretical and mathematical standpoint.

\subsection{Design Principles of Cell-Fate Decision Making Systems}
In biological systems, unlike classical or quantum computing, there is no distinct separation between hardware and software. Here, molecules act as the information-processing units (we note that this term is not without controversy), information carriers, and energy sources. Dennis Bray’s term “wetware” captures this unique nature well~\cite{Bray:2009aa}. What is more, cellular information systems are highly interconnected, lacking the isolation we see in traditional electronics~\cite{Lang:2014ia, McMahon:2015hz}.

\par
Nevertheless cells manage to be (mostly) remarkably precise in their decision making, certainly at a level sufficient to give rise to 70 or so trillion cells in an adult human, all stemming back to a single fertilized egg cell. Design principles provide a valuable framework for understanding these complex processes.
\par
To define ``design principle" \cite{Andrews2024,Poyatos2024,lynch2024evolutionary} we need to go to mathematical models, $m$, described below in \ref{sec:workdef}; here design principles are characteristics shared by all mathematical models that exhibit a particular behavior or trait. This concept, influenced by engineering and synthetic biology, has gained traction in developmental biology. Foundational work by Turing, Wolpert, and many others addressed similar ideas regarding the core principles that drive biological organization and its emergence during development~\cite{Green:2015hs, Gunawardena:2014jx}.


\begin{figure}[htb]
    \centering
    \includegraphics[width=0.90\linewidth]{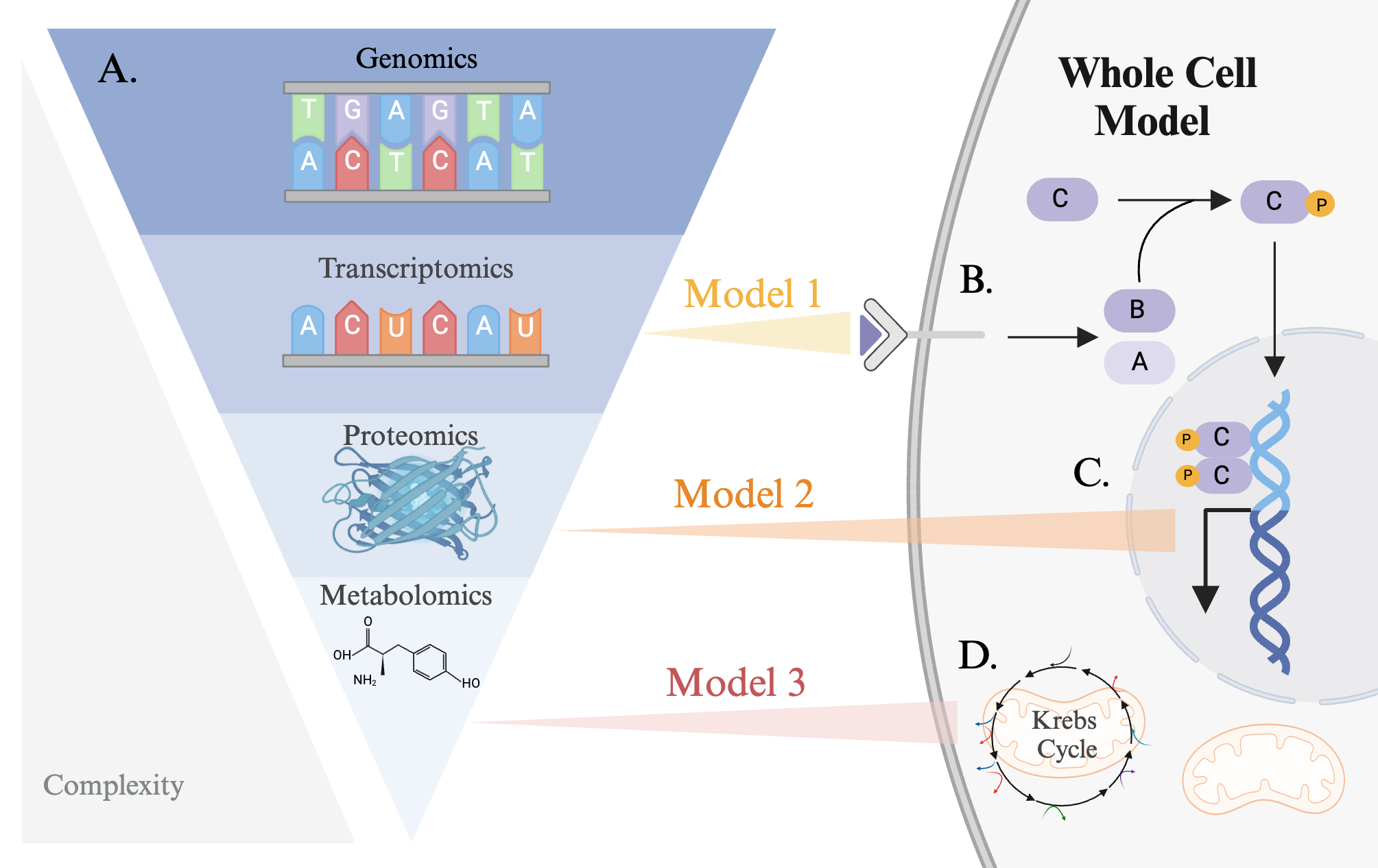}
    \caption{ Whole-cell modeling requires data from many layers of the omics pyramid (A) and integration of models which encode for the dynamics of subsystems (B, C, D). Multi-omics data varies in biological resolution and complexity (A); which are often inversely proportionate. Many computational methods model a single or subset of the microscopic systems within a cell (\textit{Model 1, 2, 3}). In order to integrate such subsystems into a biological relevant whole-cell model, their interactions and relationships must be known. As shown in this figure, \textit{Model 1} uses transcriptomic data to model signaling pathways (B) who's output regulates transcription (C); modeled with proteomics data in \textit{Model 2}. \textit{Model 3} uses metabolic data to model the Krebs cycle which is necessary for the synthesis of amino acids. This figure highlights the need for whole-cell models to follow the flow of information that drives cell-fate decisions. This figure was created in BioRender.com.}
    \label{fig:OmicsModels}
\end{figure}



Some design principles are easy to explain and reason about:
\begin{description}
    \item[Bistability:] A system achieves bistability--two or more stable states--only if a positive feedback loop is part of its underlying dynamics~\cite{Brandman:2005ci,Angeli:2004p25227}.
    \item[Biochemical oscillations:] All biochemical oscillators are driven by negative feedback, often with a time delay, to maintain cycles~\cite{Novak:2008,Zhou:2023}. 
   \item[Protein switch:] 
    These arise from proteins having multiple structural states with distinct functions, which shift in response to signals~\cite{Leon:2016gm,Alberstein:2022}. 
\end{description}

For most other biological phenomena, design principles may exist but are often subtle and complex~\cite{araujo2021}. We discuss four additional design principles relevant to cell-fate decision making in the following sections. These examples demonstrate two key points: (1) design principles are useful conceptual tools for analyzing biological systems, and (2) while they do not apply universally, when present, they can have a level of subtlety and complexity that needs to be captured at the relevant level of detail.

\subsubsection{Cellular compartments and multistability} 

Multistability is crucial for cells to adopt and maintain distinct fates. The compartmentalization of key signaling proteins within cellular structures, such as the cytoplasm and nucleus, creates spatial organization that supports this process. In eukaryotic cells, many signaling proteins actively shuttle between the cytoplasm, where they are activated through processes like phosphorylation, and the nucleus, where they can trigger transcriptional responses, before returning to the cytoplasm. This cyclical cytoplasm $\longrightarrow$ nucleus $\longrightarrow$ cytoplasm shuttling acts as a positive feedback loop that promotes multistability~\cite{Harrington:2013wv}. 
By compartmentalizing signaling pathways, cells can better process and retain information, enabling them to stabilize their state even amidst fluctuating signals. For example, MAPK proteins, which move between these compartments, have been shown to influence stem cell differentiation in vivo~\cite{Michailovici:2014hv}, highlighting that compartmentalized signaling is a key mechanism for establishing and maintaining cell-fates. This design principle was extracted by extrapolating the mathematical analysis  of positive feedback systems to more general systems \cite{FerrellJr:2012fe}

\subsubsection{Robust perfect adaptation} 
Robust perfect adaptation (RPA) is a phenomenon in which a stimulus triggers an initial response, often seen as an increase in the concentration of certain molecules, which then returns to baseline levels, even while the stimulus continues.  
This adaptive response allows cells to reset their signaling activity and avoid overstimulation, a crucial feature for maintaining stable cell states in the face of persistent signals.  Aruajo and colleagues \cite{Araujo:2018ce,Araujo:2023aa} were able to develop precise algebraic criteria to define the network structures that can achieve RPA. There are only two core network configurations that are capable of producing this adaptation, meaning that all biological networks exhibiting RPA must adopt one of these designs.  
This insight into RPA’s design principles was achieved through abstraction, which allowed the identification of underlying structural requirements without exhaustive trial-and-error. In the context of cell-fate decision making, RPA provides cells with the ability to respond selectively to new signals while maintaining a stable fate, making it an essential signal-processing mechanism in both biological and technological systems.

\subsubsection{Toggle switch behavior} 
Toggle switches allow cells to make binary decisions, often essential for specifying a cell-fate among competing options. Bistability, the ability to adopt one of two stable states, enables cells to lock into a specific fate when exposed to certain signals. Stochasticity is one of the hallmarks of many toggle switch architectures.  Noise can help push the system between states, and many designs are only bistable in a stochastic regime. In development, toggle switches help cells decide between pathways, like becoming either a neuron or a glial cell. This principle is also seen in microbial population dynamics, where toggle switches provide population-level heterogeneity, supporting bet-hedging strategies that allow populations to survive in changing environments~\cite{Kobayashi:2011dha}. 
Efforts to identify the design principles of toggle switches have relied on a combination of modeling and statistical model selection. However, there remain many different designs that can exhibit toggle switch behavior~\cite{Leon:2016gm}. This diversity suggests that toggle switches may represent a class of systems with so many possible implementations that it is difficult to define a single set of design principles~\cite{Barnes:2011hh,Leon:2016gm}.


\subsubsection{Turing patterns}
The final example we discuss here are Turing patterns, defined by Turing to be a purely chemical process by which stable spatial patterns can emerge. It is one of the fundamental patterning mechanisms in developmental biology, alongside positional information, rule-based mechanisms, and phase separation. Arguably positional information and Turing mechanisms have received the greatest attention \cite{Scholes:2019fq,Marcon:2016ia,Zheng:2016gc,leyshon2021,sharpe2019}, but were taken up primarily by the biological and mathematical communities, respectively. 
Over the past decade, however, there has been a convergence of these perspectives, with both mechanisms now widely recognized as working together to create the developmental diversity observed in nature~\cite{sharpe2019,Green:2015hs}.
\par
Turing's and later work summarized the design principles of these patterning mechanisms as: ``local activation and lateral inhibition" \cite{Turing:1952gh,Green:2015hs}. In this process, a molecule 
$A$ in one cell increases through positive feedback, diffuses to neighboring cells, and activates 
$A$ in those cells as well. At the same time, it also activates a second molecule $B$, which diffuses and inhibits the production of $A$ in neighboring cells. This interplay of activation and inhibition enables stable spatial patterns to emerge across cell populations.  


\subsection{Working Definitions for Cell States, Types, and Fates}
\label{sec:workdef}
We distinguish pragmatically between cell states, cell types, and cell fates. Here we make no attempt to further qualify these working definitions, which we believe will ultimately be necessary.
\subsubsection{Cell state}
We use this to refer to the molecular census and structural arrangement of all the molecules comprising a cell. In practical terms, not everything will be experimentally accessible. In current applications in mammalian systems, including stem cell systems, we have access to mRNA measurements at the level of individual cells. We currently do not have the ability to sample protein abundances, epigenetic states, metabolome concentrations, and more, at the sufficient scale and quality. As a result, current Cell Atlas projects provide only a partial view of cell states. 
\par
Imaging data offers a complementary perspective by revealing cell morphology and the spatial organisation of structures like organelles and molecular complexes. Combined with molecular census data, this allows us to build a more integrated view of cell states. 
\par
Many definitions of cell states overlook or minimize the influence of environmental feedback, which crucially determines the cellular physiology.  To address this limitation, we might consider defining a ``renormalized" cell state, encompassing both the molecular inventory and structural arrangement within a cell, along with interactions with its environment.

\subsubsection{Cell type} A cell type can be usefully defined as an ensemble of cells that, from a physiological or biological perspective, can be considered as equivalent. In mathematical terms this could, for example, correspond to region in cellular phase space that, under suitable thermodynamic conditions, is enriched for experimentally observable cells.
\par
Cell types are traditionally defined using morphology or cellular phenotypes. The extent to which these classical definitions of cell types align with molecular signatures remains an open question, one that we will revisit throughout this review. This question is also closely linked to understanding the sources and propagation of molecular noise within cells
\par
We can usefully differentiate between fully differentiated cell types and cell types that have some differentiation and proliferation potential. The latter include all stem and progenitor cell types. The former include all cells which no longer undergo a cell cycle, and where the only possible fate is cell death.

\subsubsection{Cell-fate} This term refers to the cell type that can be reached from an initial cell type, with an inherent temporal aspect, representing a path (or set of paths) through cellular phase space between cell types. Along this path, individual cells can be observed moving in one or both directions.

\par
Important questions include when is a cell's fate determined at the molecular level versus the phenotypic or morphological level. Similarly important questions include how many potential fates or paths are open to a given cell type; which cell types can be reached from which others; whether cell-fates are reversible, and if they are, is the same path through cellular phase space traveled in both directions.  
\par

\subsection{Cell-Fate Decisions}
The set of molecular computations and transformations of cellular states by which a cell ``decides" (or is ``forced") to follow a certain path are referred to as cell-fate decision making processes. 

\subsubsection{Biological descriptions of cell-fate decisions}
We use the term ``cell-fate decision" as a shorthand for a complex set of processes,  implying that cells possess some level of information-processing and computational ability, as well as agency: they respond and react to internal and external signals and cues and change their behavior accordingly. There are philosophical discussions to be had on this, we are sure, but we eschew these here.
\par

\subsubsection{Mathematical descriptions of cell-fate decisions}
To clarify the concepts of cell state, cell type and cell-fate developed above, we can adopt a mathematical perspective \cite{Moris:2016jt,Guillemin:2020jk}. Let $X$ denote the cell state, encompassing the abundances, and arrangements of all molecular components within the cell. We can define a suitable phase space, $\Omega_X$, in which $X$ resides,
\[
X \in \Omega_X \subseteq \mathbb{R}_0^{+n}.
\]
The temporal evolution of $X$ is described in terms of a stochastic differential equation (SDE),
\begin{equation}
    dX = \overset{\text{\footnotesize\color{blue}Deterministic Dynamics}}{\overbrace{\color{black} f(X;\theta,t)dt}}+ \overset{\text{\footnotesize\color{blue}Stochastic Dynamics}}{\overbrace{\color{black}g(X;\theta, t)dW_t}}
+\overset{\text{\footnotesize\color{blue}External Influences}}{\overbrace{\color{black}h(X;\Psi,t)dt}}.
\label{eq:SDE}
\end{equation}
Here $f(X;\theta,t)$ represents the deterministic dynamics and $g(X;\theta,t)$ the stochastic dynamics, where $W_t$ is a Wiener process capturing the inherent noise~\cite{coomer2021a}. The term $h(X;\Psi,t)$ describes the (assumed to be deterministic -- this can be relaxed) influences from the environment (e.g. a change of growth-medium).  Parameters $\theta$ define the model, while $\Psi$ governs the environment. A cautionary note for readers with experience in dealing with SDEs: in chemical or molecular processes the functional forms of $f(X;\theta,t)$ and $g(X;\theta,t)$ are tightly linked; hence the noise depends explicitly on the system's state, $X$. We therefore often refer to both terms jointly as the model 
$m$, which summarizes our understanding and hypotheses about the underlying processes in a biological system, here focusing on cell-fate decision making. 
\par
This model $m$ captures cell-fate dynamics and is often associated with Waddington's epigenetic landscape~\cite{Huang:2020fv,rand2021,coomer2021a,saez2021a}. 
These models range from simple, qualitative ``back-of-the-envelope" types to more comprehensive frameworks. They may be designed to align with experimental data~\cite{camacho-aguilar2021, saez2021a, Liu:2024aa} or to explore broad phenomenological principles~\cite{Corson:2012bb, Corson:2017ew, rand2021}. At the most detailed level, a whole-cell model, termed a "CellMap" by Sydney Brenner~\cite{Brenner:2010}, provides mechanistic, quantitative, and testable insights into cellular processes.

\par
While we focus here on SDE models of the form in Eqn.~\eqref{eq:SDE}, other mechanistic modeling frameworks can provide valuable perspectives. These include Boolean and generalized logical networks, stochastic Petri nets, agent-based models, and interacting spin-systems. Much of what we discuss here can be applied, with minor adjustments, to these other modeling frameworks. What they share is a foundational mechanistic structure that enables us to test and refine our understanding of the underlying mechanisms driving cellular behavior.
\par
Mathematical models enable us to explore fundamental questions in cell-fate decision making systems such as:
\begin{enumerate}
    \item How many cell types exist?
    \item How heterogeneous can cells of the same type be? Or in other words, how many states and how varied can these states be within a single cell type?
    \item How stable are cell types?
    \item What causes a cell to change its type and make a cell-fate decision?
\end{enumerate}
We cannot answer these questions without a tight integration of experiment, observation, and theoretical analysis.

\section{The Experimental Signatures Of Cell-Fate Decision Making Systems}
We live in a data rich world -- but there never seems to be sufficient data. We also live in a hypotheses rich world: our understanding of almost all complex biological processes and systems is still in its infancy and as a result there are many {\em a priori} plausible but mutually contradictory hypotheses about the organisation, function, and dynamics of cellular systems \cite{Clevers:2017fu,Rafelski2023}.  Currently, we lack sufficient high-quality data to thoroughly test, refine, and either validate or invalidate these hypotheses, concepts, and models related to cell-fate decision making systems~\cite{Glauche2021, stumpf2021}.
A key challenge is that testing our models often requires carefully designed experiments that yield data capable of distinguishing between different models or alternative hypotheses.

\par
In this section, we review the current experimental frameworks for probing cell-fate decision making systems, focusing on data aspects we believe will remain valuable in the near future. Instead of detailed comparisons of specific methods, we emphasize three critical issues impacting cell-fate research. First, current data is often incomplete and affected by experimental noise, which must be factored into our analyses. Second, environmental conditions, cell-cell interactions, and the cell cycle all significantly influence cell fate and can complicate inference from experimental data~\cite{Rich:2024,Guillemin:2020tm}.  Third, most current data lacks time resolution, as we are unable to monitor the same cell over time at the necessary scale and detail. We will revisit these challenges throughout the following discussion. 

\subsection{The Genomic Context}
We distinguish between two facets of the genome: first the inheritable material that is passed down through generations, subject to mutation and recombination; and second, the genome's dynamic organisation within the cell, including the 3D arrangement of chromosomes, which largely determines which genes are accessible for transcription into mRNA and subsequent translation into protein.

\subsubsection{The heritable genome}
We now have the DNA sequences of many species, and for several, we also have high-resolution data on genomic diversity, revealing genetic variability within populations. Evolutionary and comparative genomics, supported by bioinformatics tools, enable us to annotate and interpret these genomes, especially for newly sequenced organisms.
\par 
In the context of cell-fate decisions, the heritable genome plays a lesser role, except when comparing different species, such as embryonic development across mammals or varying phenotypes in yeast species, where even closely related species can exhibit very different phenotypes. Heritable genomic elements—such as gene sequences, regulatory regions, and arrangements—shape many molecular and phenotypic differences, influencing factors like gene activation timing, mRNA processing, and protein isoform diversity. However, not all sequence-level differences are shaped by selection pressures~\cite{Lynch:2006aa}, and some aspects, like the full diversity of protein isoforms, remain largely speculative~\cite{Aebersold:2018aa}. 
\par


\subsubsection{The dynamic genome}
Since the first draft of the human genome was published, our understanding of genome content and the diverse functions of non-coding DNA has transformed. We now recognise, though still incompletely, that the genome is regulated by a complex epigenetic machinery that actively opens and closes DNA sequences, modifying access to coding regions~\cite{Sarkies:2020}.
\par
These epigenetic mechanisms, while reversible, are sensitive to environmental factors and influence which genes are accessible for transcription, thereby affecting cell fate.These include DNA methylation and demethylation, histone modifications, and chromatin remodeling. Methylation is the addition of a methyl group (CH$_3$) to DNA (or proteins), often suppressing gene expression by limiting transcriptional access, playing a crucial role in maintaining or altering cell states.  Histones are the proteins around which DNA is wrapped and the tightness of this wrapping also leads to decreased accessibility of DNA to the transcriptional machinery. Chromatin remodellining adjusts the structure of chromatin to reveal or hide specific DNA sequences, directly impacting which genes are active in determining cell identity.
\par
A raft of experimental techniques are available to study these epigenetic and dynamic aspects of the genome and its {\em in vivo} geometry. ATAC-seq is used to understand patterns of DNA methylation and to analyze chromatin remodeling.  Histone modifications are assessed through CHiP-seq (Chromatin immunoprecipitation) and similar assays. Hi-C (chromsome conformation capture) is widely used to assess the 3D structure of the genome, but other methods, such as genome architecture mapping and related approaches give a more direct representation of 3D genome structure, including contacts between chromosomes. Real-time methods like live-cell imaging and single-molecule tracking operate at a lower throughput compared to other techniques. Unlike most methods, they allow us to monitor the same cells over time, though they are typically limited to tracking a small number of molecules. Collectively, these tools help us connect the physical and dynamic organisation of the genome to its functional role in cell fate decisions, enhancing our ability to map and interpret complex epigenetic landscapes.

\subsection{Gene and Protein Expression}
Most current analyses of cell-fate decision making systems are based on transcriptomic data. We can measure mRNA content in single cells and gather transcriptome-wide data for tens of thousands of cells~\cite{Regev:2018wo, Lahnemann:2020ch, Gorin:2023}. Single-cell mRNA-seq has led to the development of numerous statistical and computational tools, which have been comprehensively reviewed elsewhere. Here, we focus on key limitations: the incomplete nature of this data, the influence of confounding factors, and the lack of time-course data from direct observations, all of which have also been discussed in prior work~\cite{Gorin:2022, Gorin:2023}.
\par 
The central dogma of molecular biology describes the flow of information from genetic instructions to functional molecules: 
\begin{equation}
\text{DNA} \longrightarrow \text{mRNA} \longrightarrow \text{protein}.
\label{eq:centraldogma}
\end{equation}
Each step involves complex processes. As we have discussed above, the ``expressability" of DNA depends on numerous factors that regulate which genes can be transcribed. The amount of mRNA in a cell reflects a balance of production and degradation, as well as other regulatory mechanisms that finely tune mRNA levels and activity. Proteins are similarly regulated through degradation pathways, such as the ubiquitin-proteasome system, which plays a major role in protein turnover. Beyond simple degradation, the proteasome also engages in protein splicing, as seen with the non-ubiquitin-dependent 20S proteasome~\cite{Liepe:2016ku}. Crucially, a protein's activity is often controlled by post-translational modifications, which depend on signaling networks that respond to physiological and environmental cues, adjusting the activities of kinases, phosphatases, transcription factors, and other key molecules~\cite{Komorowski:2013ic}.

\par
Currently, we can probe mRNA at scale, but we often rely on simplified (and often linear) assumptions to link mRNA levels to active protein levels. Despite this limitation, many findings on cell-fate decision making in embryonic development, hematopoiesis, and tissue homeostasis are likely robust and will be further validated as we incorporate additional data types. However, analyses that aim to fully define cell types or reconstruct gene regulatory networks underscore the need for a more comprehensive understanding of all cellular molecules. Our view of “incompleteness” in data is slightly broader than most. While it’s true that transcriptomic data alone may leave gaps, experimental biases and limitations further contribute to an incomplete picture, even within the mRNA level.

 The primary factors complicating our analyses include, above all, the cell cycle. A range of tools are available to detect genes and their mRNA levels in relation to cell-cycle stages. Cell-cycle progression is regulated by key proteins, such as cyclin-dependent kinases, and we have an increasingly detailed understanding of cell-cycle checkpoints and their role in cell-fate decisions, although most insights remain at the protein level.

Most single-cell mRNA and emerging single-cell protein assays are destructive, meaning cells must be lysed to be analysed. This prevents true time-course studies of individual cells, so we often rely on computational methods, such as pseudo-time analysis, to infer temporal order among observed cells~\cite{Wang:2019ew, Pham:2023}. Additionally, explicit mathematical models of cell ensembles and their trajectories over time provide insights into temporal—and, hopefully, causal—relationships among molecular processes in cells undergoing fate changes.
  
\subsection{Metabolic Signatures and Determinants of Cell-Fate}
Metabolomics seeks to provide a detailed profile of the metabolic processes occurring within a cell~\cite{Rabinowitz:2012bi}. Metabolism operates at a faster rate than processes like gene regulation and was historically one of the first cellular processes amenable to mathematical analysis. It is linked to key clinical phenotypes in humans and is particularly important in microbiology, especially in biotechnological applications involving bacterial or single-celled eukaryotic organisms. In developmental biology, however, metabolic studies have largely focused on the effects of environmental changes on an embryo or the roles of nutrition and drug metabolism in health and disease.

A growing number of single-cell metabolomics techniques are becoming available~\cite{Hrovatin2023, Ahl2020, Rappez2021, Luzia:2024}, often developed with a focus on drug metabolism—a process that, in many cases, also involves cell-fate transitions. While single-cell technologies in microbial metabolomics are still developing, there is a clear understanding of the importance of integrating multiple data types. Mechanistic models play a crucial role in achieving this integration from the outset, and we discuss this in more detail in Section~$5$.

\subsection{Molecular Interaction and Regulatory Networks} 
\label{ssec:network}
Our understanding of cell-fate decision making has moved far beyond viewing genes as isolated entities. Instead the analysis of biochemical, interaction, and regulatory networks is now routinely central to analysis of cell-fate decision making systems. In mathematical terms a network, $\mathcal{N}$ is the union of a set of vertices/nodes (e.g. metabolites, proteins, or genes), $\mathcal{V}$ and a set of edges/links (chemical reactions, physical or regulatory interactions) between these vertices, $\mathcal{E}$. We then write, $\mathcal{N} = ( \mathcal{E},\mathcal{E})$. Networks have been used in mathematics, physics, and computer science, but also the social sciences.  
\par
The predominant networks in the cell biology literature are:
\begin{description}
\item[Metabolic networks] capture the biochemistry and the set of biochemical reactions happening inside a cell, as well as the influx and efflux of metabolites; here $\mathcal{N}$ are the metabolites, and $\mathcal{E}$ are the reactions connecting metabolites.
\item[Protein-protein interaction networks] provide a summary of the physical interactions between pairs or groups of proteins. Often this can include reactions that occur during signal transduction. $\mathcal{N}$ is the set of all proteins in an organism, and $\mathcal{E}$ denotes the set of  the reactions connecting metabolites.
\item[Gene regulation networks] aim to describe the complete set of transcriptional regulatory interactions and relationships; now $\mathcal{N}$ is the set of all genes, and $\mathcal{E}$ contains regulatory relationships between pairs of genes. 
\end{description}
 While each of these networks was initially studied independently, researchers now recognize the value of examining them together to capture their interdependencies.
\par
 We probably understand metabolic networks (MN)  best: the pioneering efforts of early biochemists, and arguably, the early mathematical work describing (bio)chemical and catalytic reactions and processes, have meant that we have a good understanding of many of the key metabolic processes. This is helped by the fact that the biochemistry is widely shared across the tree of life, and we have been able to apply lessons learned in one organism, say yeast, or pigeon muscle tissue, to other organisms. 
 \par
Protein-protein interaction networks (PIN) and gene regulatory networks (GRN) are, by comparison, both less well characterized, but also show greater dependence on the species than  MNs. To reconstruct PINs we have to rely on error-prone experimental assays; and GRNs we typically try to infer from gene expression data. All three networks are condition-dependent and dynamic to a degree that makes inference and analysis challenging. After a surge of interest in simple network models in the 1990s-2000s, the field now takes a more nuanced view, recognizing the complex temporal and conditional dependencies within these systems. Increasingly, we are moving from simple graph models to hypergraphs, where edges can connect multiple nodes simultaneously~\cite{Muller:2022,Zhang:2023,battiston2021a}. Here edges can have more than two incident nodes. For example, the first phosphorylation of Erk \cite{Filippi:2016gs}, 
\[
\textrm{Erk} +\textrm{P} \longrightarrow \textrm{Erk-P}
\]
 would be represented by a single edge with three nodes, Erk, P, and Erk-P.
 \par
Networks and hypergraphs thus provide a  visual representation of cellular processes, serve as an organizational tool for tracking molecules and interactions, and lay the groundwork for more mechanistic models of cell behavior.
 
\subsection{Cellular Mechanics} 
The molecular components of a cell interact not only with each other but also with their environment, and cells are influenced by various physical forces, including mechanical forces, chemical adhesion forces between cells, and electrostatic interactions within the cell.

Mechanical forces can push and pull cells, activating cellular responses—a process seen even in embryos~\cite{Ichbiag:2023, Fabrefes:2024}. The movement of cells within tissues is tightly coordinated by morphogen gradients, which guide these mechanical behaviors. However, studying these interactions is challenging, as we currently lack the ability to combine large-scale molecular assays and mechanical measurements in a single experiment. While we can apply and measure mechanical stress and use live imaging to observe responses, advances in spatial transcriptomics (and eventually spatial proteomics) will improve data collection. Nonetheless, integrating these different data types will remain a significant challenge.
\par

\section{Modeling Cell-Fate Decision Making Systems} 
\label{sec:modeling}
The processes and the data modalities that we have outlined above highlight a set of challenges and opportunities that are amenable to modeling and sophisticated data analysis. We suggest here that mathematical analysis of mechanistic models of our hypotheses will allow us to integrate data into an interpretable and quantitative framework \cite{Huang:2018cw} that will allow us to improve our understanding of biological processes.

\subsection{Data-Driven Analysis}
In this section we focus primarily on single-cell transcriptomic data. Especially in the context of single transcriptomics there have been huge advances in the analysis of single-cell data. We will primarily focus on mechanistic perspectives in the next section, but starting with networks of gene regulation. 

\subsubsection{Descriptive analysis}
We refer to \cite{Lahnemann:2020ch} for a recent review of descriptive methods for single-cell analysis. But we like to issue a note of caution: the high dimensionality, (tens of) thousands of genes across (tens of) thousands of cells has given rise to a suite of dimensionality reduction techniques. Arguably their popularity is tied to our preference to visualize data and this is only possible in low dimension. Techniques such as t-SNE and uMap are producing deceptively pretty plots and it can be tempting to apply quantitative interpretations to such visualizations, but this should be avoided as the purposes of analysis and visualization of high-dimensional data diverge and can be incompatible. 

\subsubsection{Network inference}
Networks, as discussed above, offer us the opportunity to study interacting sets of genes/proteins and gain some systems-level insights. Single-cell mRNA has given rise to a new set of network inference methods. They take the matrix of gene $\times$ cell gene expression measurements and seek to identify sets or pairs of genes that are statistically informative about each other's expression levels. Network inference has been a notoriously challenging problem in systems biology \cite{stumpf2021a,Pratapa:2020bt,Wells:2019gq}. At the core sits the large-$p$-small-$N$ problem, i.e. the fact that we have more potential hypotheses of the presence of pairwise interactions, than measurements $N$. Single-cell data with its abundances has overcome much of this.
\par
Two types of approaches have been successful: approaches that are robust to noise and which can capture non-linear dependencies \cite{Chan:2017cc}; and flexible statistical or machine learning approaches, especially if combined with additional data \cite{Petralia:2015}. The most promising methods still have appreciable error rates but serve as interpretative tools or hypotheses-generation methods for further testing. Crucially these methods, especially for carefully designed experiments, allow us to infer networks corresponding to different cell types or developmental stages.  Capturing the complex temporal dependencies among cells (using e.g. geometric perspectives) allows us even to generate cell specific networks \cite{Wang:2021}. This field is still in its infancy, but will likely result in more detailed testable hypotheses about the workings of biological processes.
  
\subsection{Hypothesis-Driven Analysis}
In Eqn.~\eqref{eq:SDE} we have assumed the existence of a mathematical model that describes the temporal evolution of a biological system. Below we provide an overview of recent attempts to model sub-cellular and cellular processes related to cell-fate decision making.
\subsubsection{Modeling molecular and cellular processes}
There have been mathematical approaches to cell cycle, cell signaling, and gene expression among others.  One of the recurring themes has been the origin and maintenance of cell-to-cell variability and its impact on cell-fate decision making. We distinguish between two types of noise: intrinsic noise is due to the random effects affecting molecular reactions and is captured by stochastic modeling approaches; the second term in Eqn.~\eqref{eq:SDE} aims to capture these effects. Extrinsic noise, the second component we consider refers to the variability that is due to factors that are not explicitly captured in our model. This can, for example, include the cellular environment but also cell cycle stage, differences in ribosome abundance, or mitochondrial activity between different cells \cite{Ham:2020kj}. Capturing both types of noise is important; dissecting sources of noise is, however, not always possible, though some experimental designs allow this in principle and in practice. 
\par
In addition to SDEs of the form ~\eqref{eq:SDE}, we can also employ other modeling frameworks, both exact and approximate, to simulate and analyze molecular processes. For a limited number of cases exact solutions or statements can be derived, but simulation is currently the only route for many situations.  We are getting better at describing the design principles for noise propagation and attenuation, but both further experimental and theoretical work is required.  For integrating transcriptomic and protein levels stochastic modeling can already help a lot and we can explore different alterations and refinements to the model and its dynamics, for example, by including opening and closing of chromatin, degradation of mRNA and protein, or threshold effects \cite{Ham:2024}. 
\par
\subsubsection{Modeling developmental processes}
To model developmental processes biologists and mathematicians often turn to Waddington’s epigenetic landscape \cite{Moris:2016jt,macarthur2022} -- a metaphor in which cell differentiation is visualized as a journey through a landscape of valleys and hills. Cells start as less differentiated at the top of the landscape, move downhill, and encounter branching points along the way. These branches represent cell-fate choices, with each path leading to a distinct valley corresponding to a stable cell type. This metaphor, introduced nearly a century ago, has since inspired quantitative and computational frameworks to describe key qualitative dynamics of developmental systems, such as the stability of cell types, bifurcations leading to different cell-fates, and the role of intermediate cell states. Over time, approaches to modeling the developmental landscape have evolved into two broad categories: ``gene-centric"\cite{camacho-aguilar2021,saez2021a,Liu:2024aa} and ``gene-free models"\cite{Corson:2012bb,Corson:2017ew,rand2021}, each offering distinct perspectives, advantages and disadvantages. 



\subsubsection{Gene-centric models of the developmental landscape}
Gene-centric models are typically rooted in SDEs of the form \eqref{eq:SDE} that describe the evolution of gene expression profiles over time. By presuming a network of gene interactions, these models seek to understand how individual genes and their interactions shape the developmental landscape. Specifically, the probability 
$P(X)$ of a cell’s gene expression vector $X$ is used to determine valleys (stable states) and paths (transitional states) within the so-called ``quasi-potential" landscape: 
\begin{equation}
U(X) \propto - \log( p(X)).
\label{eq:quasipotential}
\end{equation}
\par
The simple picture conceived by Waddington and analyzed by many since then, has provided valuable insights into the molecular mechanisms underlying stability and transition points within the developmental landscape. These models are effective at capturing the influence of specific gene interactions on cell behavior, which has been important for understanding the roles of critical genes in driving cell-fate choices and maintaining stable cellular states~\cite{Brackston:2018bb,Brackston:2018kw}.

As we connect this metaphor to the underlying mathematics and biological data, however, certain limitations surface.
For example, these models typically rely on the a gradient assumption, where the drift component of the underlying SDE is derived from the gradient of a potential function. The system’s dynamics are thus guided by an ``energy-minimizing” behavior, with cells tending to move ``downhill” in this landscape to reach stable states.
This gradient-based perspective has limitations, especially in biological systems with complex, oscillatory, or non-conservative behaviors. Such processes do not strictly follow energy-minimizing paths, which means that they cannot be fully captured by a gradient potential alone. Moreover, the gradient assumption is typically applied within a quasi-steady-state context, where gene expression dynamics are presumed to stabilise over time. This assumption may not hold in developmental systems that undergo active transitions, bifurcations, or shifts between cell states \cite{Saez:2022aa}. 

Stochastic dynamics, intrinsic to cellular behavior, can reshape the landscape both quantitatively and qualitatively, introducing elements such as noise-induced transitions. These can drive transitions across the landscape in a way that diverges from simple ``downhill” behavior. Differential equation models, however, typically use deterministic frameworks or assume additive noise.  Gene-centric models also overlook cell-cell interactions and spatial or temporal factors, which are critical in many developmental processes but are challenging to integrate into a fixed gene network framework.

\begin{figure}[htb]
    \centering
\includegraphics[width=0.90\linewidth]{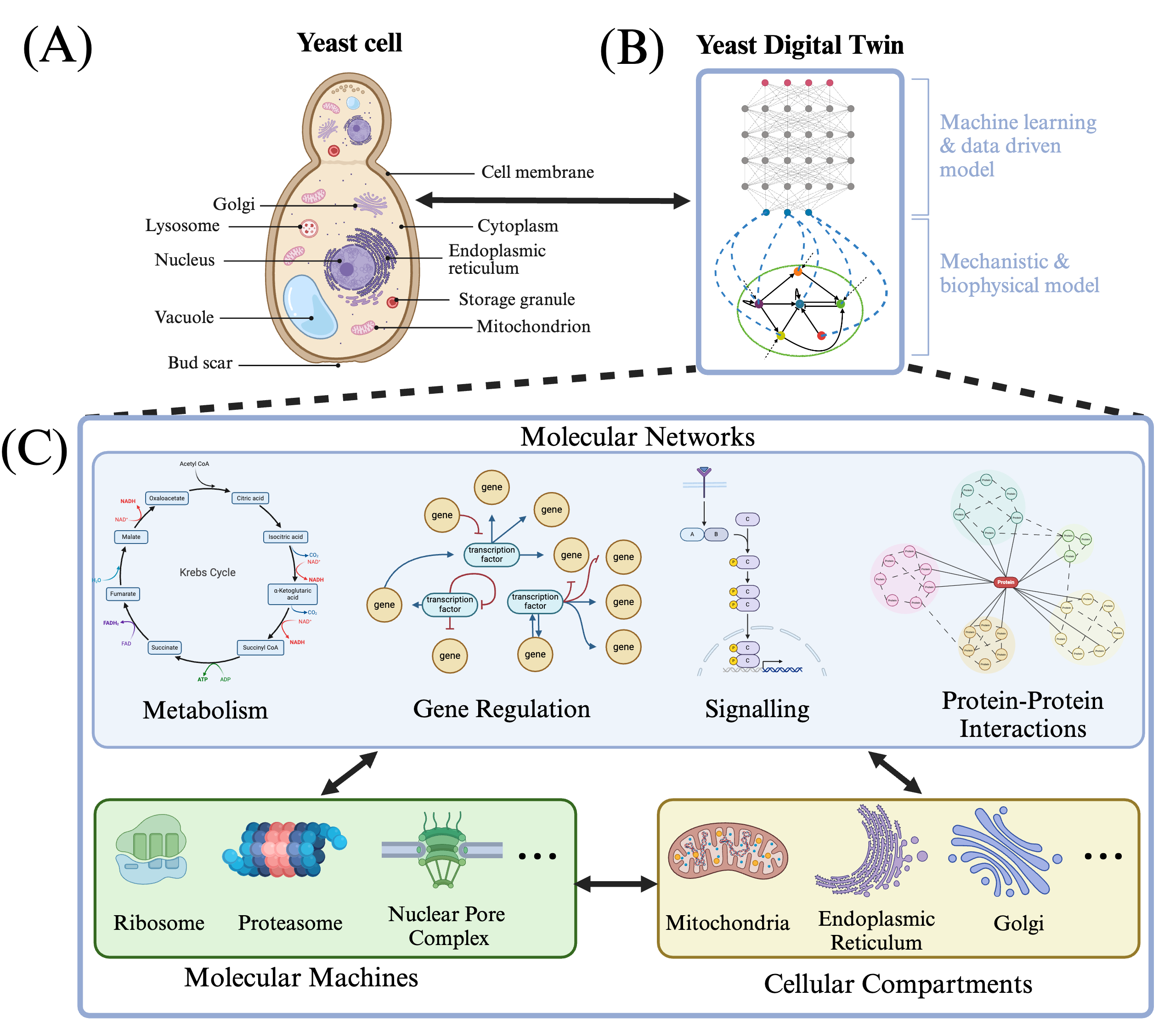}
\caption{ (A) Yeast cell with major cellular compartments highlighted. Our ambition is to provide mathematical models, or ‘digital twins’, for fungal organisms beyond {\em Saccharomyces cerevisiae}. (B) Our digital twins are hybrid models which combine the advantages of mechanistic and biophysical modeling with machine learning/AI models of less well characterized system components that can be learned from experimental data. (C) The mechanistic model is constructed from available biological information, including metabolic, gene regulation, signaling, and protein-protein interaction network data; the molecular machines that make up these networks; and the cellular compartments into which the cell is organized. This information can be curated from bioinformatics resources, public data bases, and literature (through text-mining). This figure was created in BioRender.com.}
\label{fig:digtwin}
\end{figure}

\par
\subsubsection{Gene-free approaches in developmental modeling}
In parallel to connecting models of landscapes to gene regulatory systems, recent approaches explore the structure of these landscapes without relying on specific assumptions about gene networks. This shift, pioneered in part by the work of Briscoe, Siggia, and Rand,~\cite{Corson:2012bb,Corson:2017ew,rand2021} leverages the work of renowned mathematician Ren\'e Thom, who had a keen interest in applying catastrophe theory to biological systems.  This ``gene-free" perspective diverges from traditional gene-centric models, instead focusing on the global structural features that characterize developmental processes.  These approaches seek to describe the behavior of the system as a whole, using concepts like stability points and saddle nodes to study the dynamics of cell state transitions. Rather than anchoring models in predefined gene networks, this framework captures the system’s global behavior, focusing on the topological and geometrical features that shape developmental landscapes.

Despite their innovative approach, these methods rely on the same gradient-based assumptions as gene-centric models and are thus subject to similar limitations when connecting these with data.  However, they do offer several advantages.
By not relying on specific gene interactions, gene-free approaches provide flexibility in capturing the complex and adaptive behaviors of developmental systems, accommodating phenomena that are difficult to model within a gene-centric framework.
The structural features identified through catastrophe theory, such as bifurcations and attractors, offer robust insights into cell state stability and the conditions that lead to cell-fate transitions.
Such methods lend themselves well to high-dimensional single-cell data, where the explicit modeling of all gene interactions is computationally prohibitive and often impractical due to incomplete information on gene regulatory networks.
As the field progresses, integrating these approaches with empirical data may illuminate previously uncharacterized aspects of developmental biology, such as transient cell states and non-equilibrium dynamics, offering a pathway toward comprehensive, system-level models of development.

Ultimately, combining gene-centric and gene-free models may yield a more comprehensive and nuanced understanding of developmental processes, accommodating both gene-specific interactions and broader system-level behaviors. This combined framework has the potential to capture a richer, multidimensional picture of development, opening new pathways for exploring cell-fate determination and dynamic regulatory landscapes.

\section{Hybrid Models Of Cell-Fate Decision Making Systems And Their Applications In Biotechnology }
\label{sec:hybridmodels}

Up to now we have dealt with cell-fate decision making systems in multicellular systems and predominantly in the context of developmental processes. But cell-fate decision making is not exclusive to the eukaryotic kingdom of life, nor to multicellular organisms. Prokaryotic and archaeal cells, but also single-celled eukaryotic organisms, also have to respond and adapt their behavior to different environments. Just as in the case of eukaryotes, there are decisions that are part of the normal cell physiological processes such as adapting to different environmental or internal cellular conditions.  Prokaryotes -- we focus on these for simplicity and because our experience overlaps with this domain -- exhibit bistability~\cite{Gardner2000-fg}, oscillatory behavior~\cite{Elowitz2000-ko}, a multitude of switch-like behaviors (including toggle-switches)~\cite{Gardner2000-fg}, robust perfect adaptation~\cite{Yi2000-vd}, and even Turing patterns~\cite{Turing:1952gh}; they exhibit this behavior either naturally or in synthetically engineered systems. In short, they have  considerable computational capacities, some of which we can harness and even repurpose to our advantage. 
There would be neither bread nor beer without {\em Saccharomyces cerevisiae}, but genetically engineered cells are poised to revolutionize a number of different industries, including manufacturing, across a much broader range of applications. Harnessing and repurposing microbial metabolism promises a sustainable way to produce chemicals and materials -- from fuel to food and cosmetic additives to pharmaceuticals -- that have traditionally relied on fossil fuel-based manufacturing practices.
\par

\subsection{Synthetic Biology to Control Cell-Fate Decision Making Processes} 
 Synthetic and engineering biology try to optimize and repurpose biological processes, even whole microbial organisms, for applications in bio-manufacturing and beyond. One promising avenue here is to reprogram or rewire the cell-fate decision making machinery of these biotechnologically important organisms using new genomic constructs. Understanding cell-fate decision making systems, the ability to explore their behavior, and the behavior of synthetically engineered systems {\em in silico}, promises to make this design process more efficient. The hope is that {\em in silico} investigations can triage the biological design space before {\em in vitro} or {\em in vivo} analyses, reducing the time and associated costs involved in creating new microbial strains. Early successes show that this hope seems justified~\cite{Voigt2020-bt}.
 \par
 Understanding the design principles that drive certain types of cellular behavior would greatly assist in designing appropriately performing biosynthetic systems.  As such, the search for design principles has become central to many endeavours in engineering biology, and generally proceed in one of three ways. First, we can use mathematical analysis, typically by exploring and exploiting abstractions of mathematical models that can exhibit the desired behavior. Robust perfect adaptation is one example where this is possible. Where this is not possible -- and that is for the overwhelming majority of cases -- we can either exhaustively explore model spaces when the design space is relatively small; or we can sample models statistically to explore larger model spaces. However, even this will ultimately reach a limit because of the associated computational demands. For example, previous analyses that have exploited exhaustive sampling of Turing pattern generating (TPG) mechanisms considered thousands of models (5,760 different models with three nodes, where two are allowed to diffuse). Evidently, if we want to explore parameter spaces, at least to a moderate degree, then the number of model evaluations quickly explodes. In these situations we can either make mathematical simplifications which allow analytic insights into approximate models~\cite{Marcon:2016ia,leyshon2021,Smith:2018gc}, or sample less idealized (and arguably more realistic) models~\cite{Zheng:2016gc,Scholes:2019fq}. Comparisons between different assessments of potential Turing pattern generators then depend on the nature of the systems and, for sampling-based approaches, on the depth of sampling. For example, a more comprehensive analysis where some $10^{11}$ different model/parameter combinations were sampled~\cite{Scholes:2019fq} found more TPGs than an approach which sampled some $10^8$ combinations~\cite{Zheng:2016gc}.
 \par
 A recurring problem in these studies is that the models we consider are systematically and purposefully designed to focus on what are perceived as the core mechanisms, often ignoring the wider molecular and cellular networks in which these mechanisms are embedded. While this approach can be successful, there are no {\em a priori} guarantees. When it fails, it is difficult to decide whether the issue is due to inaccuracies in the model itself, or external factors modulating cellular dynamics beyond the model's scope~\cite{Gunawardena:2014jx,Kirk:2015gj}. Such cell physiological feedback has been observed in many contexts as we have already argued above. One way to address this challenge is to make models larger. For example, in engineering, we might first model and design an airplane wing in isolation. However, before building a physical prototype, we refine this design within a comprehensive computer model of the entire plane, since airflow around an isolated wing differs from that around a full aircraft. Despite the increased computational cost, we scale the models to capture these complexities. Similarly, we believe it may be essential to develop more detailed models of cellular systems that function as ``digital twins" of real cells. This concept, dating back at least to discussions between Sydney Brenner and Francis Crick in 1967, is older than many realize.

\subsection{Whole-cell Models and Hybrid Whole-cell Models}
Constraint-based metabolic models have been widely used in the design of genetically engineered organisms and microbial strains. While often successful, models focusing solely on metabolism -- so ignoring gene regulation, signaling, cell-wall biophysics, and other biological processes --  have inherent limitations~\cite{Kirk:2015gj}. First, they model only the enzymes involved in metabolism, overlooking genes with functions outside of metabolic reactions. Second, they fail to capture the interactions between different biological processes and how these interactions shape the behavior of the system as a whole. Sydney Brenner's idea of a CellMap~\cite{Brenner:2010} is more comprehensive, and efforts towards whole-cell modeling are well underway. Particularly for bacteria, starting with {\em Mycoplasma genitalium}~\cite{Karr:2012bh} ($\approx$525 genes) and advancing to {\em E. coli}~\cite{Sun2021-li}, and for yeast, we now have powerful models that extend beyond metabolism~\cite{elsemman2022} to include cellular compartments, gene regulation, and signaling. The aim is to generate ``digital twins" of these industrially important microbes, allowing them to be studied {\em in silico} to establish cause-and-effect relationships that connect biological and physiological mechanisms across scales. This approach can ultimately guide the design of biosynthetic strains with superior performance characteristics (Fig.~\ref{fig:digtwin} A,B).   
\par
Whole-cell models are conceived to mathematically capture all {\em relevant} biomolecular processes inside a cell, Fig.~\ref{fig:digtwin}B. As we have argued above, mathematical or mechanistic models are particularly well-suited to integrating data across multiple 'omics levels to uncover mechanisms that explain the emergence of system-level behavior. However, in the short term at least, we lack the knowledge to construct these models comprehensively despite the wealth of information on metabolic, gene regulatory, and signaling processes (Fig.~\ref{fig:digtwin}C). Instead, we now have an opportunity to generate hybrid models that combine mechanistic and data-driven modeling frameworks in ways that leverage their distinct features and compensate for the deficiencies of the other. 
Machine learning can incorporate physics-based knowledge, e.g. governing equations, boundary conditions, and constraints, to resolve ill-posed problems and non-physical predictions that may arise from the naive use of ML on sparse, noisy, or biased data. Additionally, ML can benefit from mechanistic models by using them to generate synthetic data, providing a cost-effective and efficient way to supplement sparse training datasets. In turn, mechanistic models can leverage machine learning to develop surrogate models that approximate complex, computationally intensive mechanistic simulations as seen in~\cite{Gherman2023-mq}, identify system dynamics and parameters, analyse sensitivities, and quantify uncertainty.
\par
Currently, there are very few genuinely hybrid models that integrate mechanistic and data-driven approaches. More often, ML is employed to aid in the construction of mechanistic models. For example, in~\cite{Li2022-sr} the authors use deep learning to predict catalytic constants ($k_{cat}$ values) across different species which are subsequently used as inputs to enzyme-constrained genome-scale metabolic models (ecGEMs). Similarly, ML is frequently used to extract features from multi-omics data for setting flux balance analysis (FBA) constraints~\cite{Rana2020-ir}.   
In a truly hybrid model we encode our knowledge and our hypotheses in a mechanistic mathematical model, e.g. of the type in Eqn.~\eqref{eq:SDE}; but we also model other less clearly defined processes and cellular features, including uncertainty in our model, through a ML/AI model such as a deep neural network, Fig.~\ref{fig:digtwin}B. From a design perspective, these models naturally perform best when design alternatives can be represented within the mechanistic framework. 
\par
There are promising applications in strain design, for example, that demonstrate the benefits of combining metabolic modeling with ML/AI approaches, whether truly hybrid or otherwise~\cite{Sahu2021-gz,Zhang2020-hf}. Ultimately, this hybrid approach has the potential to vastly reduce the time and effort required to generate whole-cell models for different species. This has a range of different advantages. First, generating digital twins for related organisms will give us the ability to explore the complex genotype-phenotype relationships theoretically, and to address fundamental questions in evolutionary genomics from a new perspective e.g. the differences in pathogenicity among closely related fungal species. Second, it allows us to study, and even expand (using natural or synthetic strains), our repertoire of biotechnologically useful microbial systems.
\par
Complex interacting networks that define cell-fate decision making systems are now captured in rich datasets, enabling the development and integration of mechanistic and machine learning models to deepen our exploration and understanding of how cells ``make" decisions.  
\section{Summary} 
Purely descriptive analysis methods cannot describe the complex processes we observe in cell-fate decision making systems. We need to have comprehensive descriptions that allow us to integrate diverse and heterogeneous data into a single framework. Decision making is a process with an intrinsic time component but so far we have lacked the ability to observe these processes directly but have instead relied on computational processes to infer temporal dynamics. 
\par
Here we argue that CellMaps or mechanistic models of cellular behavior are essential to make sense of the large amounts of heterogeneous data available to us. Among many important functional insights they are also pivotal to distill design principles of cellular behavior. If we are able to identify such design principles we will have a framework to reason about cell-fate decision making processes, especially if we are able to consider cell fate decision making systems from across the tree of life. Integrating and interpreting functional and conceptual drivers of cell-fate decision making between species, and across experimental systems, will become a greater priority.
\par
We are on the cusp of being able to affect cell-fate decision making processes {\em in vivo} and need to build up the ethical and societal understanding and license to do so. Here, too, CellMaps and whole cell models will be one important tool to assess safety, reliability, and social acceptability prior to any {\em in vitro} and {\em in vivo} interventions.

\section*{Disclosure Statement}
MAC and MPHS are co-founders and shareholders of Cell Bauhaus. 

\section*{Acknowledgments}
TEW \& MPHS acknowledge funding through an Australian Research Council Laureate Fellowship to MPHS (FL220100005). LH acknowledges funding throught the ARC Centre of Excellence for the Mathematical Analysis of Cellular Systems. We thank the members of the Theoretical Systems Biology group for lively discussions and the stimulating atmosphere that have been the setting in which this review became possible. 
%
\bibliographystyle{ar-style3.bst}
\bibliography{ar.bib}

\end{document}